\documentstyle[12pt]{article}

\newskip\humongous \humongous=0pt plus 1000pt minus 1000pt

\newif\ifdtup

\catcode`\@=11

\@addtoreset{equation}{section}
\def\theequation{\thesection\arabic{equation}}

\def\@normalsize{\@setsize\normalsize{15pt}\xiipt\@xiipt
\abovedisplayskip 14pt plus3pt minus3pt%
\belowdisplayskip \abovedisplayskip
\abovedisplayshortskip \z@ plus3pt%
\belowdisplayshortskip 7pt plus3.5pt minus0pt}

\def\small{\@setsize\small{13.6pt}\xipt\@xipt
\abovedisplayskip 13pt plus3pt minus3pt%
\belowdisplayskip \abovedisplayskip
\abovedisplayshortskip \z@ plus3pt%
\belowdisplayshortskip 7pt plus3.5pt minus0pt
\def\@listi{\parsep 4.5pt plus 2pt minus 1pt
     \itemsep \parsep
     \topsep 9pt plus 3pt minus 3pt}}

\relax

\catcode`@=12

\catcode`\@=11

\def\section{\@startsection{section}{1}{\z@}{3.5ex plus 1ex minus
   .2ex}{2.3ex plus .2ex}{\large\bf}}

\def\thesection{\arabic{section}.}

\def\appendix{\setcounter{section}{0}
 \def\thesection{Appendix \Alph{section}:}
 \def\theequation{\Alph{section}.\arabic{equation}}}
\begin{document}

\begin{titlepage}
\begin{center}
{\Large
Level-Crossing in the Instanton-Anti-Instanton Valley}
\end{center}

\vspace{1em}
\begin{center}
{\large
Riccardo Guida, Kenichi Konishi and Nicodemo Magnoli }
\end{center}

\vspace{1em}
\begin{center}
{\it Dipartimento di Fisica -- Universit\`a di Genova\\
     Istituto Nazionale di Fisica Nucleare -- sez. di Genova\\
     Via Dodecaneso, 33 -- 16146 Genova (Italy)\\
     E-mail: Decnet 32655; Bitnet KONISHI@GENOVA.INFN.IT\\}
\end{center}

\vspace{7em}
{\bf ABSTRACT:}
We study the level crossing of the fermion system described by the
 euclidean Dirac Hamiltonian in the valley background.
 One chiral fermion level is shown to cross twice the zero value in the
case  of
well-separated  instanton-anti-instanton background.
 Below a critical separation, however, level crossings
are  absent.  The phenomenon can be interpreted
as the transition  to a gauge field configuration of purely perturbative
nature,
 below a  critical instanton-anti-instanton
separation.   In the context of high-energy
electroweak interactions,  our findings seem to definitely invalidate
some  optimistic
argument concerning the  observability of baryon number violation
 based on the use of the optical theorem in conjunction with
the valley fields.
\vspace{2em}
\begin{flushleft}
GEF-Th-14/1993~~~~~~~~~~~~~~~~~~~~~~~~~~~~~~~~~~~~~~~~~~~~~~~~~~~~~~~~
{}~~~~~~~~~ June 1993 \end{flushleft}
\end{titlepage}

%%%%%%%%%%%%%%%% latex definitions
\newcommand{\beq}{\begin{equation}}
\newcommand{\eeq}{\end{equation}}
\newcommand{\bea}{\begin{eqnarray}}
\newcommand{\eea}{\end{eqnarray}}
\newcommand{\beas}{\begin{eqnarray*}}
\newcommand{\eeas}{\end{eqnarray*}}
\newcommand{\defi}{\stackrel{\rm def}{=}}
\newcommand{\non}{\nonumber}

%%%%%%%%%%%%%%%%%%%%%%%%%%%%%%%%%% definitions

\def\dirac{{\cal D}}
\def\dplus{{\cal D_{+}}}
\def\dminus{{\cal D_{-}}}
\def\dbar{\bar{D}}
\def\H{\cal{H}}
\def\de{\partial}
\def\si{\sigma}
\def\sb{{\bar \sigma}}
\def\rn{{\bf R}^n}
\def\r4{{\bf R}^4}
\def\s4{{\bf S}^4}
\def\ker{\hbox{\rm ker}}
\def\dim{\hbox{\rm dim}}
\def\sup{\hbox{\rm sup}}
\def\inf{\hbox{\rm inf}}
\def\infi{\infty}
\def\nrm{\parallel}
\def\nrmi{\parallel_\infty}
\def\teo{\noindent{\bf Theorem}\ }
\def\tt{\tilde T}
\def\st{\tilde S}
\def\om{\Omega}
\def\Tr{ \hbox{\rm tr}}

\section{Introduction.}
The purpose of the present paper is to gain a deeper insight in the
behaviour of chiral fermions in non-Abelian gauge theories  in the background
of the so-called valley fields (i.e., a particular class of
gauge backgrounds of instanton-anti-instanton type),
 by analyzing the spectral flow of the
corresponding euclidean Dirac Hamiltonian.

The interest in this issue arose from  the problem of the observability
of the anomaly-induced fermion-number violation at high energies
in the standard electroweak theory,
from the study of the vacuum structure in QCD,
and also from the questions related to the large-order behaviour of
perturbation theory in non-Abelian
gauge theories in general.

The problems involved, in particular the one
 related to unitarity and chiral anomaly,
have been studied by two of the present authors in Ref.\cite{guida}.
 It was shown there
that the
 forward elastic four-fermion amplitude computed in the valley backgrounds
indeed satisfies unitarity by having an anomalous piece, proportional
to the products of the standard fermion zero modes,\footnote{i.e,
of pure instanton or of pure anti-instanton background. Recall that no
four-dimensional zero modes exist in the valley
background.\protect\cite{guida}}
  in spite of the fact that
the relevant gauge field is a topologically (globally) trivial one.
Furthermore, it was found that
 such an anomalous contribution is there as long as the instanton
anti-instanton  ($i - a $) separation is large enough.

On the other hand, at small $i - a $  separation the amplitude reduces to
a perturbative one.  In fact, on the basis of the behaviour of the integrated
topological density,
\beq C(x_4) = - \int_{-\infi}^{x_4}\int d^3x {g^2\over 16 \pi^2}
\Tr\,F_{\mu \nu}{\tilde F}_{\mu \nu} ={\cal N}_{CS}(x_4)- {\cal
N}_{CS}(-\infi),
 \label{top}      \eeq
where
${\cal N}_{CS}(x_4)\equiv - {g^2\over 16 \pi^2}\int d^3x \epsilon^{4ijk}
\Tr (F_{ij} A_k - {2\over 3}A_i A_j A_k) $
 is the Chern-Simons number,
the same authors argued that such a transition to purely perturbative
 amplitude should  takes place abruptly at around the unit $i - a $
separation,  $ R/\rho \simeq 1 $.
 However, the studies
of Ref.\cite{guida}
 left unresolved the question exactly at which value of the $i - a $
separation this occurs, and did not quite
 clarify the nature of this transition.

 In this note we close that gap
 by studying the zero level crossing (spectral flow) of the chiral fermion
system described by the euclidean
 Dirac Hamiltonian in the valley background.
Our numerical analysis clearly
reveals a transition in the behaviour of the spectral flow at
a  critical value of separation, $R = R_c \simeq \sqrt{4\over 3} \rho$,
$\rho$ being the common size of the instantons.
 For $R>R_c$
we find two level crossings,
located near the positions of instanton and anti-instanton,
while for $R<R_c$  no level crosses zero.

\section{Spectral flow}
Let us consider an $SU(2)$ gauge theory with  $N_F$ massless left-handed
 doublets.  We consider in particular the fermions in
a fixed external gauge field,
\bea
A_{\mu}^{(valley)}& =&-{i\over g}(\sigma_{\mu} {\bar \sigma_{\nu}}
-\delta_{\mu \nu} )\,[\, {(x-x_a)_{\nu} \over (x-x_a)^2 +\rho^2 }   \non \\
& &
+
{(x-x_i)_{\nu} \rho^2 \over (x-x_i)^2 ((x-x_i)^2 +\rho^2) }
 + {(x-x_i +y)_{\nu} \over
(x-x_i + y )^2  } -  {(x-x_i)_{\nu} \over
(x-x_i)^2  }\,]
\label{valley} \eea
where
\beq
 y^{\mu} = -R^{\mu}/(z-1);\,\,
 z = (R^2 + 2 \rho^2 + \sqrt{R^4 + 4\rho^2 R^2}) /2\rho^2;\,\,
R^{\mu} = (x_i - x_a)^{\mu}. \label{parameters} \eeq
$x_i^{\mu}$ and $x_a^{\mu}$ are the centers of the instanton and
 anti-instanton, $\rho$ is their (common) size.  The gauge fields of Eq.(\ref
{valley})
 has been  termed the valley (or streamline) trajectory
(\cite{Yung}).

$N_F$ must be an even number in order for the theory to be well defined.
  These fermions can be  rearranged (in an $SU(2)$ gauge
theory) to  ${N_F \over 2}$ Dirac
fermions by appropriately combining  pairs of  left-handed fermions and their
righthanded anti-fermions; the resulting Dirac fermions are coupled
 vectorially to the gauge boson. In this case the axial charge will
be anomalous, corresponding to the  fermion number violation
 of the original model.

For simplicity of notation, let us restrict below
 to $N_F=2$, equivalent to the case of a
single Dirac
doublet. The generalisation to the case with $N_F$ chiral fermions is
straightforward.
 The euclidean  Dirac  operator is $i\gamma_\mu D_\mu$,
where $D_\mu=(\de-igA)_\mu$ is the usual covariant derivative.
In the Lorentz gauge
 of Eq.(\ref{valley}) the gauge field can be expressed as:
\beq \label{field}
A_\mu =-{i\over g} (\si _\mu \sb _\nu -\delta_{\mu \nu})
\, F_\nu = \bar{\eta}_{\mu\nu}^a F_\nu \si ^a\eeq
\beq
F_\mu(x)\equiv {1\over 2}\,\de_\mu\log L(x);\;\;\;\;\;\;
L(x) \equiv {(x-x_a)^2+\rho^2\over (x-x_i)^2+\rho^2}\,(x-x_i+y)^2.
\label{elle}\eeq
The instanton and anti-instanton location will be chosen at:
$x_i=(-{R\over 2},0,0,0)$ and $x_a=({R\over 2},0,0,0)$; as a consequence
the vector $y$ above (\ref{parameters}) is in the time direction
$y=(y,0,0,0)$.

Decomposing
$
i\gamma _\mu D_\mu=i\gamma_0 (D_0 + \H)
$
as
$$
\cal{H}= \pmatrix {H_{+} & 0\cr
                 0 &H_{-}},$$
and passing to a noncovariant formalism,
\footnote{Usually the spectral flow is studied in the axial gauge where
$A_0=0$. The Hamiltonian in that gauge
would be related to (\ref{hamiltonian}) by a time-dependent gauge
transformation of spatial components
of the gauge field that however  preserve the spectrum of the Hamiltonian and
correspondingly spectral flow.}
 we find:
\beq
 H_{+}=-H_{-}=H
\label{condition}\eeq
 with
\beq
H=+i\vec{\si}\cdot\vec{\nabla}
-\vec{\si}\wedge \vec{\tau}\cdot\vec{F}-F_0 \vec{\si}\cdot\vec{\tau}
\label{hamiltonian}\eeq
 where
$$x_\mu\equiv(t,\vec{x});\;\;\;\vec{F}=\vec{x}f(r,t);\;\;\; F_0=F_0(r,t).$$
$r=\sqrt{{\vec{x}}^2}$;  $\vec{\si} $ are spin operators, and
$\vec{\tau } $ are isospin one.

Clearly $\H$ is a self-adjoint operator with entirely essential spectrum,
that is,  eventual (normalisable) eigenvalues
are imbedded into the continuum, which
covers  the real axis of energies.

To have a picture of well separated levels, some of which crossing zero
at certain values of the euclidean time\footnote{Note that
 $\H$ depends explicitly on $t$.}, one
must consider the Dirac operator on a compactified space, $S^4$ (sphere)
or $T^4$ (torus), but one must make sure that the compactification preserves
the zero modes present in the continuum theory.
We shall bypass the whole
 problem, just studying the eventual zero modes
of $\H$ for some values of $t$ as the  signal  of level crossing, and
 disregarding the  more complex problem
of the  full analysis of time evolution of levels.

 The  condition (\ref{condition})
implies that zero modes always appear in pair of different chirality.
We assume that for each chirality  zero modes are non degenerate, so that
we must look for singlets of the total angular momentum (which is a symmetry
of $\H$),  $\vec{J}=\vec{x}\wedge -i\vec{\nabla}+\vec{\si}+\vec{\tau}$.
As noted in \cite{guida} the most general form of
the singlet is:
\beq
\eta(\vec{x},t)=\si _2 S(r,t)-i \vec{x}\vec{\si} \si _2 T(r,t).
\eeq

The equation $H\eta =0$ then reads:
\footnote{ We take into account the fact
 that the only spatial vector is $\vec{x}$.}
\bea
3 F_0 S-3T-\vec{x}\vec{\nabla}T+2 \vec{F} \vec{x} T&=&0, \\
-F_0 r^2 T +\vec{x}\vec{\nabla} S+2\vec{F} S&=&0.
\eea

Making the substitution
\beq
S={1\over L}\st, \qquad
T={L\over r^3} \tt,
\eeq
 we get a  simplified system:
\beq \label{system}
\tt '(r;t)=3F_0 {r^2 \over L^{2}} \st ; \qquad
\st '(r;t)=F_0 {L^2\over r^2} \tt
,\eeq
where the prime means differentiation with respect to $r$ and the
time $t$ appears as a
 parameter.
Normalisability of the solution implies:
\beq \label{normalisation}
\infi >\int_0^{\infi}d\!r r^2  |S|^2=\int_0^{\infi}d\!r {r^2\over L^2}
|\st|^2 ;\qquad
\infi >\int_0^{\infi}d\!r r^4 |T|^2=\int_0^{\infi}d\!r {L^2\over r^2}
|\tt|^2.\eeq
Recall that with our choice of parameters we have:
\bea\label{functions}
L &\equiv& {t_a^2+r^2+\rho^2\over t_i^2+r^2+\rho^2}\,(t_y^2+r^2)\\
F_0&\equiv&{t_a\over  t_a^2+r^2+\rho^2}-{t_i\over  t_i^2+r^2+\rho^2}
+{t_y\over  t_y^2+r^2+\rho^2}
\eea
where
$
t_a\equiv t+{R\over 2};\;\;t_i\equiv t-{R\over 2};\;\; t_y\equiv t-
{R\over 2}+y.$
We want to know for which range of the parameter $R$ and for which values
of $t$ - if any - (the other
parameter $\rho$ just fixes the scale) the system Eq.(\ref{system}) has
a normalisable solution.
Clearly, normalisability
enforces the following initial condition (at a given $t$ )
\footnote{Note the qualitative
 change in the behaviour of $L$ which occurs at $t_y=0$.}:
\bea
&&\st (0)=1;\;\; \tt (0)=0;\;\;\hbox{\rm if}\;\; t_y\neq 0\label{initial1}\\
&&\st (0)=0;\;\; \tt (0)=1;\;\;\hbox{\rm if}\;\; t_y = 0. \label{initial2}
\eea

The  problem turns out to be
 too hard  to be
 treated exactly: it just suffices to note that the system is equivalent to
a second order linear equation for $\tt$  with $12$ regular fuchsian points!
We thus proceeded to a numerical method for finding solutions of the problem
(\ref{system}), (\ref{normalisation}) and (\ref{initial1}) or (\ref{initial2}).

In passing we recall \cite{Kiskis}
 that for the case of a pure anti-instanton
(or an instanton)
a normalisable solution can be found easily. For instance,
in the case of an  anti-instanton (centered at $x_{\mu}=0$)
 the gauge field has the
same form as (\ref{field}), with
$
F_\nu={x_\nu /( x^2+\rho^2)}$
so that
\beq F_0={t\over t^2+r^2+\rho^2}, \qquad
     L=t^2+r^2+\rho^2.\eeq
The normalisation condition (\ref{normalisation}) requires in this case
that
\beq
\st (0)=1;\;\; \tt (0)=0.\eeq
It is clear that for $t=0$ (corresponding to the time position of
the anti-instanton)
$F_0=0$ for all $r$ and the system decouples.  It follows
that
$ \st(r)=1;\;\; \tt(r)=0$,\,\, i.e.,
$$S(r,0) = {1\over r^2 + \rho^2};\;\;\;\;T(r,0) = 0 $$
is the desired  normalisable solution.  Furthermore, it can be
shown that $t=0$ is the only value
of the parameter for which such a solution exists.

\section{Numerical solution of the system.}

Solving the system by a power series in $r$ (method of Frobenius), we find
the asymptotic behaviour of two independent solutions of
(\ref{system}), which is reported
\footnote{Disregarding the numerical coefficients.}
 in Table 1 for  $r\sim 0$ and in Table 2 for  $r\rightarrow \infi $.
We treated separately the special cases,
 $t={R\over 2}-y$ (corresponding to $t_y=0$),
 $t = t_0$ (where $t_0$ is by definition a zero of $F_0(r=0)$ ),
$t=t_1$ (where $t_1=-{R\over 2}-y$).

Clearly  the solution satisfying (\ref{initial1}) or
(\ref{initial2}) can be easily selected
near $r=0$; the problem is
how this solution evolves at $r\rightarrow \infi$.
To see this an accurate numerical analysis is required. We have made such a
numerical analysis, passing to a compactified variable
$p=r/(1+r)$, and using as initial conditions at a point very near
\footnote{Note that the coefficients of the system are singular at $r=0$, so
that numerical evolution must start a little after zero, typically $r=10^{-9}$
(in $\rho$ units).}
$p=0$ (i.e. $r=0$) the values of $\st$ and $\tt$ which
are  obtained as a power-series approximation of  the adequate
solution of Table 1. The common normalisation  of $\st$ and $\tt$
  at this point can be
arbitrarily set to unity.
Then we studied whether the solution evolves towards $r = \infi$
in such a way that the normalisation
condition (\ref{normalisation}) is satisfied.

First, we find that for  $t_y=0$ the solution
satisfying the initial condition (\ref{initial2}), is not normalisable.

For  $t_y\neq 0$,  we ask whether  $\tt (r=\infi)=0$,
i.e., whether
 the solution of the initial condition problem is  normalisable.
In Fig. 1a  $\tt (r=\infi)$ is plotted as a  function of $t$  for
 $R=10$: it is seen that the curve intersects  zero twice.
Note that the two values of
 $t$  at which the solution
is normalisable are very close to the (time) positions of the instanton and
anti-instanton.

   The singularity in Fig. 1a is located at $t_y=0$ ($t\sim 5.1$), and is
due to the fact that there are no solutions of the system for this value
of $t$ that
 satisfy the condition (\ref{initial1}):
see Table 1.

%In Figure 2 (and Figure 3) the behaviour in $r$ of solutions corresponding to
%different values of time near the zero crossings is reported ($R=10$).

We find furthermore that the situation is similar for  all values of
 the instanton-anti-instanton separation above a critical value, i.e.,
 $R>R_c$.
(See also Fig. 1b  where $R=1.5$).  The critical separation turns out to be
 \footnote{The value $\sqrt{4\over 3}$ is a
fit to our numerical result.}
\beq{R_c\over \rho} \simeq 1.15470....\simeq \sqrt{{4\over 3}}.\eeq
At   $R= R_c$ the two  level-crossing points  coalesce
into one.

Below the critical distance, $R < R_c$,  on the contrary,
no values of $\,t\,$ are found for
 which  the system Eq.(\ref{system}) has a normalisable solution. (See
Fig. 1c that reports the situation just below the threshold, for $R=1.0$).

For large instanton anti-instanton
 separations, our calculations thus  indicates
 the  presence of level crossings
near the positions of instanton and anti-instanton centers, that seems to
reproduce locally the situation mentioned above
for a single (anti)-instanton field.
This means that the lowest  level $H_{+}$ (in a
 compactified space), for instance,
starts from a  positive value at  $t= -\infi $ and crosses
zero to  negative,   then crosses  zero back to positive, and
returns to the original value at $t =\infi$. A level of opposite chirality
also crosses zero twice but in the specular manner with respect to
the $E=0$ line.

At the separation below the critical value $R_c$ there is  a qualitative
 change of behaviour: no levels cross zero.
 The valley field
thus appears to lose the topologically non-trivial  aspect and degenerate into
a field configuration of purely perturbative nature. (At $R=0$ the valley field
Eq.(\ref{valley}) is indeed a gauge transform of $A_{\mu} = 0 $.) In other
words, the
instanton-anti-instanton pair "melt" at $R\simeq \sqrt{(4/3)}\rho$.

Before closing this section we observe
that the  critical value of $R$, $R = R_c\simeq\sqrt{4\over 3}\rho $
corresponds to the value of the conformally
invariant variable $z$ (see Eq.(\ref{parameters})),
  very close to $z = 3$.   And this corresponds \footnote{We thanks
 P.Provero for pointing  this out to us.} to
 the maximum
of the integrated topological density,
equal to  $C(x_4=0)={1\over 2}$ (see
Eq. (4.3) of  \cite{guida}
 for the explicit expression in terms of $z$).
 The situation is very similar to what happens in the
Schwinger  model on the cylinder (QED on  $R^1 \otimes S^1$)
 (see \cite{shifman}), where the level
crossings disappear at the same value of the analogous topological quantity.
It is interesting that this value of  $C(x_4=0)$
  just corresponds to
the top of the hill separating the two adjacent minima of the gauge
field action.

\section{Discussion}

We  propose here a physical interpretation of the mathematical analysis
made above, on the level crossing of the euclidean Dirac Hamiltonian.

Consider the (euclidean) time evolution of a given state from $t=-\infi$
 to $t=\infi$.  For definiteness first consider  the vacuum-to-vacuum
transition. In the
first quantized
approach, the  wave
function is given   at $t=-\infi$  by the Dirac sea
with all negative energy levels filled.  Then we follow its (euclidean)
 time evolution in the
Schr\" odinger picture.
Suppose \footnote{The adiabatic approximation suppresses the unavoidable
perturbative creation of particle-anti-particle pairs of
same chirality (hence with no net chirality change)
that is present for every time-dependent background,
allowing  us to concentrate on the
 nonperturbative phenomenon related to the level crossings.}
that the
time evolution   is  adiabatic.  This means that an
eigenfunction of $H(t)$ with eigenvalue $E_n(t)$ evolves at a successive
 instant $t'$
into an eigenfunction of $H(t')$
with eigenvalue $E_n(t')$, corresponding to  the original one.
 If one level crosses $0$ during the
 evolution, for example from negative to positive,
 the state we started with will   no longer be  the vacuum of the theory
 at the instant $t'$
but corresponds to the state with one positive energy level filled, that is
to a one-particle state (where "particle" means here "quantum of energy", see
\cite{labonte} for a possible, more accurate, definition).
 Analogous situation for creation of anti-particles,
corresponding to empty negative levels (holes) with respect to the new
 vacuum.

In Section 3 we found that in the valley background with
$R > R_c$ one left level crosses  zero twice (and so does a right level).
 This suggests \footnote{We assume on  physical grounds
 that  in the valley the level crossing near the  instanton location is such
that a left (right) positive (negative) level crosses zero downwards
(upwards) as in the case of  a single instanton field; and an
opposite situation  near the anti-instanton.}
 that during the time evolution every quantum
state acquires an extra lefthanded fermion
and an antiparticle of the righthanded fermion
 near  the anti-instanton position, hence with
the net chirality change, $-2$.\footnote{A more precise statement
is  $\Delta Q_5 \equiv \{Q_R - Q_L\}|_{t_a+} - \{Q_R - Q_L\}|_{t_a-}=-2
 $: whether this means the creation of two lefthanded particles, the
annihilation of two righthanded particles, or e.g., the creation of one left
particle and the annihilation of a right particle, depends on the particular
process considered and is a matter of no importance.}
 An analogous net chirality change $+2$ occurs  near the instanton center.
In the case of vacuum-to-vacuum transition this would mean the propagation
of two lefthanded fermions between
 the instanton centers, signalling the instability of the vacuum.
The vacuum-to-vacuum amplitude would get an imaginary part, corresponding to
 intermediate states with $Q_5 =  2$ or $ Q_5 = -2.$

Although the physical interpretation of the level-crossing similar to this
 is commonly used (\cite{gross},\cite{shifman}),  a rigorous universally
accepted interpretation seems to be lacking in mathematical physics
 literature
\cite{particle}:
we do not pretend to go beyond our simple picture here.

In the case of elastic four fermion amplitude (two fermion-
two fermion transition) as the one considered in Ref.\cite{guida}
(relevant to the problem of high energy fermion number violation in the
electroweak theory),
the presence of the level-crossings for the backgrounds with well-separated
instanton  and anti-instanton ($R>R_c$) precisely corresponds to the
anomalous term in the amplitude,\cite{guida}
  with an imaginary part associated with
 intermediate states without the initial fermions.

The disappearance of the level crossings at and below
the critical $i-a$ separation
$R= R_c \simeq \sqrt{4\over 3} \rho$, on the other hand, means that
 the anomalous,
nonperturbative contribution to the amplitude  is absent in these backgrounds.
This confirms  the idea that the valley background Eq.(\ref{valley}) ceases to
be topologically significant long before $R$ reaches $0$ where $A_{\mu}$ is the
vacuum field.\footnote {As is well known, the
form of the valley field is not unique, and depends on the choice of
the so-called weight function. However they all share the common
property of reducing to the vacuum field at $R=0$. We therefore expect
the presence of a critical separation for any such field, below which
the level crossings disappear.}

It is important, though it might appear surprising at first sight,
 that the gauge background,
Eq.(\ref{valley}),
which is perfectly smooth as a function of $R$ and $\rho$, leads to a
discontinuous physical result at $R = R_c$.\footnote{This conclusion is not
really surprising.  As is well known from the theory of phase
transitions, physical results in
 a system with an infinite number
 of degrees of freedom can have  a nonanalytic dependence on the parameters
of the system, even if the Hamiltonian is analytic in these.
  A relativistic system of fermions  we are
concerned with here, is just such a system.}
   This conclusion is also  corroborated by (and independently implied by)
 the known
 behaviour of
the action \cite{Khoze}  and  of the integrated
 topological density \cite{guida}  $C(0)$, Eq.(\ref{top}),  as functions
 of $R$.  Indeed,  both of them behave as
$\sim R^2$    for  $R/\rho \ll 1$,
which is clearly and simply a reflection of the perturbative, quadratic
fluctuations  around $A_{\mu}=0$.\footnote {This is evident in the
"clever" gauge used in \protect\cite{guida} in which $A_{\mu}\propto R \,$ for
small $R$. }

In the physics context of Tev-region electroweak interactions, the outcome
 of our analysis is that the valley configuration at $R < R_c$, hence with
action  $S < S_c ={16 \pi^2\over g^2}(0.5960...), $  has no relation at all
to
the instanton-induced fermion number violating processes.  The argument
made in some literature \cite{Khoze}
 that  such processes become unsuppressed, on the basis of the behaviour
of the valley field at $R=0$,  thus seems to be unfounded.\footnote {We
agree on this point with a conjecture made in \protect\cite{maggiore}.}
  The valley field
is relevant only up to
  an energy where
 the cross section  is still exponentially suppressed
($\sigma_c = \exp - {16 \pi^2\over g^2}(0.3184...)$).

As for the problem of the QCD vacuum,  a
 series of works in Ref. \cite{liquid}
have led to the picture of  the physical vacuum
(with broken chiral symmetry ) as a sort of
instanton liquid,  with the average  instanton-anti-instanton distance
about
three  times their mean sizes.  Our result of  the minimum
separation $R_c/ \rho \simeq
\sqrt{4/3} $ for the instanton anti-instanton
 configuration to be non-perturbative,
 seems to support their assumptions.

Eventual implications of our work to the study of
 the large-order behaviour of perturbative series in non-Abelian
gauge theories, are still to be worked out.

\bigskip
{\bf Acknowledgments} The authors are grateful to
S. Forte, C. Imbimbo, G. Morchio and P. Provero
for interesting discussions.

{\bf Figure Captions}

\begin{description}

\item
{Fig. 1.} Behaviour of $\tt (r=\infi )$ in function of time, for
a) $R/\rho=10$, b) $R/\rho=1.5$, c) $R/\rho=1.0$.

\end{description}

\newpage
%%%%%%%%%%%%%%%%%%%%%%%%%%%%% TABELLA 1 %%%%%%%%%%%%%%%%%%%%%%%%%%%%%%%%
%%%%%%%%%% thanks to Prof. S.Frixione for explaining me how to do this table
%%%%%%%%%% and the other one.

 \begin{table}
 \begin{center}
 \begin{tabular}[bht]{|c||c|c|c|c|}  \hline
   & $t$ generic & $t=t_0$ & $t=t_1$ & $t={R\over 2}-y$ \\
 \hline\hline
   $\st_1$
 & $ 1+r^2$
 & $ 1+r^4 $
 & $ 1+r^2$
 & $ 1+r^2$
 \\
   $\tt_1$
 & $ r^3$
 & $ r^5$
 & $ r^3$
 & $ r^{-1}$
 \\
   $\st_2$
 & $ r^{-1}$
 & $ r$
 & $ r^{-1}$
 & $ r^3$
 \\
   $\tt_2$
 & $1+r^2$
 & $1+r^6$
 & $1+r^2$
 & $1+r^2$
\\
  \hline
 \end{tabular}
 \vskip .3cm
 \end{center}
 {
 \leftskip 1cm
 \rightskip 1cm
 {\bf Table 1}: Behaviour of the two linearly indipendent
 solutions  near  $r= 0$.

 }
 \end{table}
%%%%%%%%%%%%%%%%%%%%%%%%%%%%%%%%%%%%%%%%%%%%%%%%%%%%%%%%%%%%%%%%%%%%%%%

%%%%%%%%%%%%%%%%%%%%%%%%%%%%% TABELLA 2 %%%%%%%%%%%%%%%%%%%%%%%%%%%%%%%%
%%%%%%%%%% thanks to Prof. S.Frixione for explaining me how to do this table
%%%%%%%%%% and the other one.

 \begin{table}
 \begin{center}
 \begin{tabular}[bht]{|c||c|c|c|c|}  \hline
   & $t$ generic & $t=t_0$ & $t=t_1$ & $t={R\over 2}-y$ \\
 \hline\hline
  $\st_3$
 & $ r$
 & $ r $
 & $ r^{-1}$
 & $ r$
 \\
 $\tt_3$
 & $ 1+r^{-2}$
 & $ 1+r^{-2}$
 & $ 1+r^{-6}$
 & $ 1+r^{-2}$
 \\
 $\st_4$
 & $ 1+r^{-2}$
 & $ 1+r^{-2}$
 & $ 1+r^{-6}$
 & $ 1$
 \\
 $\tt_4$
 & $r^{-3}$
 & $r^{-3}$
 & $r^{-5}$
 & $r^{-3}$
 \\ \hline
 \end{tabular}
 \vskip .3cm
 \end{center}
 {
 \leftskip 1cm
 \rightskip 1cm
 {\bf Table 2}: Behaviour of the two linearly indipendent
solutions, at  $r\rightarrow\infi$.

 }
 \end{table}
%%%%%%%%%%%%%%%%%%%%%%%%%%%%%%%%%%%%%%%%%%%%%%%%%%%%%%%%%%%%%%%%%%%%%%%

\end{document}